\documentclass[aps,prb,reprint,superscriptaddress,floatfix]{revtex4-2}

\usepackage{amssymb}
\usepackage{graphicx}   
\usepackage{color}      
\usepackage{hyperref}   

\def\lco{La$_2$CuO$_4$}
\def\lsco{La$_{2-x}$Sr$_x$CuO$_4$}
\def\lbco{La$_{2-x}$Ba$_x$CuO$_4$}

\def\ybco{YBa$_2$Cu$_3$O$_{6+x}$}

\def\bscco{Bi$_2$Sr$_2$CaCu$_2$O$_{8+\delta}$}

\def\newr{\color{black}}

\begin{document}

\title{From non-metal to strange metal at the stripe-percolation transition in \lsco}

\author{J. M. Tranquada}
\email{jtran@bnl.gov}
\affiliation{Condensed Matter Physics and Materials Science Division, Brookhaven National Laboratory, Upton, New York 11973-5000, USA}
\author{P. M. Lozano}
\altaffiliation{Current address: Advanced Photon Source, Argonne National Laboratory, Argonne, Illinois 60439, USA}
\affiliation{Condensed Matter Physics and Materials Science Division, Brookhaven National Laboratory, Upton, New York 11973-5000, USA}
\affiliation{Department of Physics and Astronomy, Stony Brook University, Stony Brook, NY 11794-3800, USA}
\author{Juntao Yao}
\affiliation{Condensed Matter Physics and Materials Science Division, Brookhaven National Laboratory, Upton, New York 11973-5000, USA}
\affiliation{Department of Material Science \&\ Chemical Engineering, Stony Brook University, Stony Brook, NY 11794-3800, USA}
\author{G. D. Gu}
\affiliation{Condensed Matter Physics and Materials Science Division, Brookhaven National Laboratory, Upton, New York 11973-5000, USA}
\author{Qiang Li}
\email{liqiang@bnl.gov}
\affiliation{Condensed Matter Physics and Materials Science Division, Brookhaven National Laboratory, Upton, New York 11973-5000, USA}
\affiliation{Department of Physics and Astronomy, Stony Brook University, Stony Brook, NY 11794-3800, USA}

\date{\today} 

\begin{abstract}
The nature of the normal state of cuprate superconductors continues to stimulate considerable speculation.  Of particular interest has been the linear temperature dependence of the in-plane resistivity in the low-temperature limit, which violates the prediction for a Fermi liquid.  We present measurements of anisotropic resistivity in \lsco\ that confirm the strange-metal behavior for crystals with {\newr doped-hole concentration $p = x > p^\ast \sim 0.19$} and contrast with the non-metallic behavior for {\newr $p<p^\ast$}.  We propose that the changes at $p^\ast$ are associated with a first-order transition from doped Mott insulator to conventional metal; the transition appears as a crossover due to intrinsic dopant disorder.  We consider results from the literature that support this picture; in particular, we present a simulation of the impact of the disorder on the first-order transition and the doping dependence of stripe correlations.  Below $p^\ast$, the strong electronic interactions result in charge and spin stripe correlations that percolate across the CuO$_2$ planes; above $p^\ast$, residual stripe correlations are restricted to isolated puddles.  We suggest that the $T$-linear resistivity results from scattering of quasiparticles from antiferromagnetic spin fluctuations within the correlated puddles.  This is a modest effect compared to the case at {\newr $p<p^\ast$}, where there data suggest that there are no coherent quasiparticles in the normal state.
\end{abstract}

\maketitle

\section{Introduction}

There are a variety of normal-state properties of hole-doped cuprate superconductor compounds that deviate from expectations of Fermi liquid theory.  For example, there are the anomalous temperature dependences of properties such as magnetic susceptibility, in-plane resistivity $\rho_{ab}$, the Hall effect, and various spectroscopic features observed especially in underdoped cuprates that have been discussed commonly in terms of pseudogap phenomena \cite{batl96,timu99,tall01,norm05,lee06,hufn08,tail10,keim15,kord15}.  The crossover temperature $T^\ast$ associated with the pseudogap behaviors decreases as the doped-hole concentration $p$ increases, extrapolating to zero at $p^\ast\sim0.19$ \cite{tall01}.  In a cuprate with a lower superconducting transition temperature ($T_c$), such as \lsco\ (LSCO), where accessible magnetic field strengths can suppress the superconducting order, another anomaly appears for $p\sim p^\ast$ \cite{daou09b,coop09}; this is strange-metal behavior, in which $\rho_{ab}$ varies in linear proportion to $T$ in the low-temperature limit, in violation of the Fermi-liquid prediction of $T^2$ variation \cite{varm89,ande92}.  This behavior extends over the range $p^\ast\lesssim p < p_c$ \cite{huss11,legr19,ayre21}, where $p_c\sim0.3$ is the critical point at which superconducting order disappears \cite{spiv08}.

The cause of the strange-metal behavior has continued to be a major topic of discussion \cite{phil22,varm20b,hart22,zaan19,chow22,pate23,seib21,capr22,bagg23,bane21}.  To the extent that strange metal behavior is associated with $p^\ast$, it begs the question of the underlying physics of the cuprate phase diagram.  A crucial unresolved issue concerns whether there should be a first-order transition as a function of doping as one moves from the antiferromagnetic insulator at $p=0$ to a more or less conventional metal at $p>p_c$.  Anderson argued from the beginning that the parent cuprates, such as \lco, are Mott insulators \cite{ande87} and later made the case that the superexchange mechanism underlying the antiferromagnetism is incompatible with Fermi-liquid theory \cite{ande97a}.  In response, Laughlin argued \cite{laug98} that adiabatic continuity should apply from the overdoped metallic phase through the full doping range where superconducting order exists, so that if there is a Mott insulator phase at $p=0$, there should be a first order transition at a $p$ between the antiferromagnetic state and the onset of superconductivity.  From his perspective, the pseudogap behavior could be attributed to a competing order that is compatible with conventional band theory ($d$-density-wave order in his case), with $p^\ast$ corresponding to a quantum critical point (QCP) and strange-metal behavior resulting from critical scattering \cite{laug14}; others have considered a QCP due to spin-density-wave (SDW) \cite{tail10,teix23}, charge-density-wave (CDW) \cite{cast97,tsve14,seib21,capr22,bagg23}, or pair-density-wave \cite{bane21} orders.  

\begin{figure}[t]
 \centering
    \includegraphics[width=0.95\columnwidth]{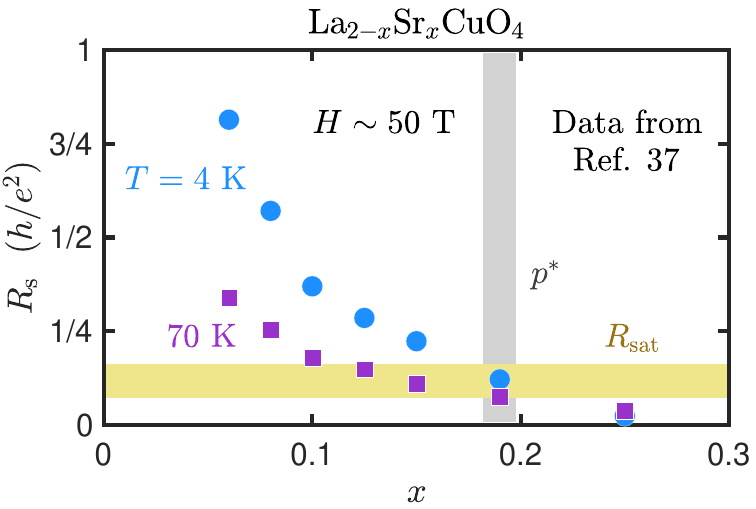}
    \caption{\label{fg:rsh} Sheet resistance, $R_{\rm s}$, in units of $h/e^2$, measured on thin films of \lsco\ in $c$-axis magnetic fields of $\sim50$~T at temperatures of 4~K (blue circles) and 70~K (violet squares) from Capara {\it et al.} \cite{capr20}.  Yellow horizontal line indicates the saturation resistance of conventional metals from Hussey {\it et al.} \cite{huss04}; gray vertical line indicates $p^\ast$ from Tallon and Loram \cite{tall01}.}
\end{figure}

Laughlin proposed that the only way to distinguish a state incompatible with Fermi-liquid theory would be to look at the behavior of $\rho_{ab}$ in the low-temperature limit of the normal state, which is obscured by superconductivity in the absence of a magnetic field.  At that time, Ando and Boebinger \cite{ando95,boeb96} had discovered that on suppressing the superconductivity in LSCO with a $c$-axis magnetic field of 60~T, both $\rho_{ab}$ and $\rho_c$ grew at low temperature as $\ln(1/T)$ for $x\lesssim0.17$; they characterized this as an insulating state \cite{boeb96}.  The results for $\rho_{ab}$ have now been reproduced by Caprara {\it et al.}\ \cite{capr20} who applied fields of $\gtrsim45$~T to LSCO thin films.   In Fig.~\ref{fg:rsh}, we show their results, measured in high field, in the form of sheet resistance, $R_{\rm s}$, at $T=4$~K and at 70~K as a function of $x$; in field, there is a rise in $\rho_{ab}$ at low temperature for {\newr $x=p\lesssim p^\ast$}.  Whether or not the resistivity truly diverges as $T\rightarrow 0$, this behavior is inconsistent with a conventional metallic state.  

For a two-dimensional (2D) metal treated by Boltzmann theory, the Mott-Ioffe-Regel limit on resistivity (electronic mean free path equals the lattice spacing \cite{ioff60}) corresponds to $h/e^2$ \cite{werm17}.  The data shown in Fig.~\ref{fg:rsh} are all below, but a significant fraction of, this limit.
For conventional metals with large electron scattering, the resistivity is empirically observed to saturate at a value of $200\pm100$~$\mu\Omega$-cm \cite{mooi73,gunn03,huss04}, which corresponds to a sheet resistance for LSCO of $R_{\rm sat} \sim 0.12h/e^2$.   
While $R_{\rm sat}$ should have some dependence on carrier density \cite{werm17}, we note the striking coincidence that the crossover from non-metallic to metallic temperature dependence at $p^\ast$ occurs where $R_s$ crosses through the typical $R_{\rm sat}$.

Such anomalous behavior provides justification for considering underdoped cuprates in terms of a model of a hole-doped Mott insulator.  Based on early computational evidence for phase separation of doped holes and antiferromagnetic spins in a doped $t$-$J$ model \cite{emer90},  Emery and Kivelson \cite{emer93} proposed that inclusion of long-range Coulomb repulsion would yield frustrated phase separation, which can take the form of charge and spin stripes \cite{low94}.  Experimental studies on LSCO and related compounds provide evidence that charge and spin stripes develop together \cite{tran95a,hunt99,savi02,birg06,crof14}. It is important to note here that, while these correlations can be described as intertwined spin density wave (SDW) and charge density wave (CDW) orders, they are a consequence of competition between antiferromagnetism driven by superexchange \cite{ande87,ande97a} and the frustrated kinetic energy of the doped holes \cite{frad15}.  They are distinct from the individual SDW or CDW orders that can develop in conventional metals and that have been considered by some \cite{tail10,cast97,seib21,capr22,bagg23}.

Given that cuprates with $p\gtrsim p_c$ appear to have Fermi-liquid character \cite{coop09,plat05,vign08,kram19}, reconciling this with the strongly-correlated behavior of underdoped samples appears to imply that there is a first-order transition from doped Mott insulator to conventional metal somewhere under the superconducting dome.  Complicating the identification of such a first-order transition is the presence of significant charge disorder.  The role of intrinsic disorder on the anticipated first-order transition in cuprates has been considered previously \cite{burg01}, and it has been noted that, even in a half-filled correlated system, disorder with percolative character can lead to a metallic response \cite{szab20}.  The picture presented here is less extreme than the latter, but anticipates the first-order transition at higher-doping than the former.

Working with others, we have recently reported a study of overdoped LSCO \cite{li22} in which we presented evidence for inhomogeneous superconductivity and made the case that the heterogeneity was a natural consequence of the random substitution of Sr$^{2+}$ ions for La$^{3+}$ together with the absence of effective screening.  In the present article, we present further measurements on our LSCO crystals.  We begin by presenting ac susceptibility measurements of the superconducting transition to demonstrate that, for each $x$, there is a single sharp bulk superconducting transition, consistent with previous reports \cite{bozo16,he16}.  Then we consider $\rho_c(T)$ and $\rho_{ab}(T)$ for compositions spanning $p^\ast$.  In particular, we demonstrate that $\rho_c(T)$ changes from insulating to metallic character as {\newr $p$} increases through $p^\ast$.  In a corresponding fashion, $\rho_{ab}(T)$ changes from flattening out near $T_c$ for {\newr $p < p^\ast$} to exhibiting strange-metal behavior for {\newr $p > p^\ast$}, consistent with previous work \cite{coop09}.

We then review experimental results in the literature that support the interpretation that $p^\ast$ corresponds to a percolative transition between the stripe-correlated phase and a conventional metallic phase.  Superconductive pairing is optimized where the stripe phase is strong \cite{tran21a}, and bulk superconductivity weakens as regions of stripe phase are limited to non-percolating puddles with overdoping \cite{li22,spiv08}.  Within this picture, it appears that the normal-state resistivity for $p>p^\ast$ should be dominated by the scattering of quasiparticles from isolated patches of strongly-correlated electrons.  This is consistent with the two charge-carrier components inferred by Ayres {\it et al.} for a variety of cuprates \cite{ayre21} and with aspects of some recent models \cite{pate23,seib21,capr22,bagg23,rice17,lee21c,ma23}, especially including recent numerical studies of the Hubbard model \cite{huan19,huan23,wu22}.

The rest of this paper is organized as follows: In Sec.~\ref{sc_exp}, we describe our experimental methods, with the characterizations (ac susceptibility and anisotropic resistivity) of our LSCO crystals presented in Sec.~\ref{sc_res}.  The interpretation of the results, and especially the nature of $p^\ast$, are described in Sec.~\ref{sc_int}.  The relevance of our results to other studies are considered in Sec.~\ref{sc_dis}, which is followed by our conclusions in Sec.~\ref{sc_con}.

\section{Experimental Methods}
\label{sc_exp}

The \lsco\ crystals used here are the same ones previously studied in Refs.~\onlinecite{li18,miao21,li22}.  They were grown at Brookhaven by the travelling-solvent floating-zone method.  Post-growth annealing in O$_2$ is described in Ref.~\onlinecite{li22}.  {\newr Based on the previous analysis, we believe that the oxygen-vacancy density is minimal and assume $p = x$.}

The ac magnetic susceptibility was measured in a 7-T Quantum Design Magnetic Properties Measurement System with a SQUID (superconducting quantum interference device) magnetometer. The measurements were conducted under a zero dc field, with an applied ac magnetic drive amplitude of 2~Oe at a frequency of 100~Hz.  Separate measurements were performed with the ac magnetic field either parallel or perpendicular to the $c$-axis.

The resistivity measurements were performed on rectangular-plate-shaped single crystals with four different doping values: $x=0.17$, 0.21, 0.25, and 0.29. Samples were oriented and cut for both in-plane and out-of-plane resistivity measurements for each doping level. The electrical contacts were done in a four-point in-line configuration, and the measurements were performed in a Quantum Design Physical Property Measurement System equipped with a 14-T superconducting magnet.

\section{Results} 
\label{sc_res}

\subsection{ac susceptibility}

\begin{figure}[t]
 \centering
    \includegraphics[width=0.95\columnwidth]{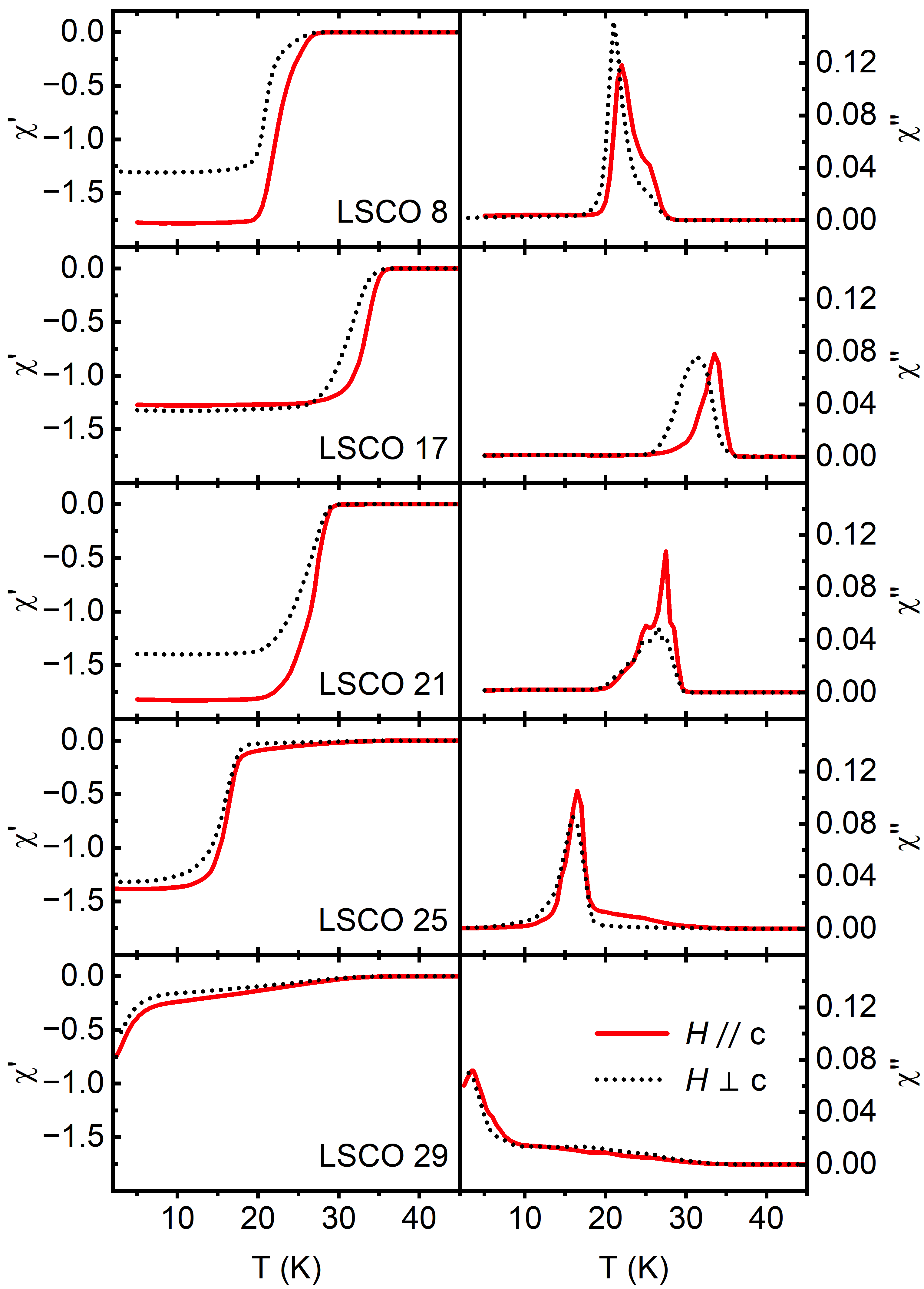}
    \caption{\label{fg:ac}  Results for the volume magnetic susceptibility, $\chi=\chi'+i\chi''$, with $\chi'$ plotted on the left and $\chi''$ on the right, obtained with magnetic field ${\bf H} || {\bf c}$ (solid red line) and ${\bf H} \perp {\bf c}$ (dotted line), for \lsco\ crystals with (from top to bottom) $x=0.08$, 0.17, 0.21, 0.25, and 0.29. {\newr We use dimensionless SI units.  Full volume shielding should yield $\chi'=-1$; data exceed that limit because of uncorrected demagnetization factors due to non-ideal sample shape \cite{osbo45}.}}
\end{figure}

Our results for the magnetic susceptibility are shown in Fig.~\ref{fg:ac}.  The temperature at the peak of $\chi''$ provides a measure of $T_c$.  As one can see, there is a single narrow peak in $\chi''$ for $x=0.17$ with ${\bf H} || {\bf c}$, which measures the superconducting transition within the CuO$_2$ planes.  The fact that the peak for ${\bf H} \perp {\bf c}$ occurs at a slightly lower temperature is consistent with the fact that the superconducting response between the planes is dependent on Josephson coupling between the planes; note that the temperature at which $\chi''$ begins to rise on cooling for both field orientations is essentially the same.  A similar anisotropy has been reported in measurements of superconducting stiffness on ring shaped crystals at the same composition \cite{sama24}.

For $x=0.25$, we see a single peak $\chi''$ at $\approx17$~K; however, $\chi''$ begins to rise from $T>30$~K.  For $x=0.29$, the peak has moved below 5~K, but the initial rise is still above 30~K.  It is of interest to compare with work on high-quality LSCO thin films across the overdoped regime \cite{bozo16}, where the superconducting transition was measured by mutual inductance at frequencies of 20--90 kHz with ${\bf H} || {\bf c}$ \cite{he16}.  In the imaginary part of the mutual inductance, there is a single narrow peak for each sample.  Similar measurements on LSCO films in an earlier study by a different group \cite{lemb11} show peaks with a bit more width and structure, closer to our results.  In any case, we argue that there is a strong qualitative similarity between our results and those for thin films \cite{he16,lemb11}.  The tails to high temperature in our results are somewhat subtle; nevertheless, they are consistent with the results from dc magnetization reported in \cite{li22}, where we made the case for an onset of granular, non-percolating superconductivity above the bulk transition for the $x=0.25$ and 0.29 crystals.  {\newr That conclusion is also consistent with the results on polycrystalline samples with multiple O$_2$ treatments \cite{taka92a}, where samples with $x=0.25$ and 0.30 showed the onset of weak-but-finite Meissner response at $T>35$~K.}

\begin{figure*}
 \centering
    \includegraphics[width=1.7\columnwidth]{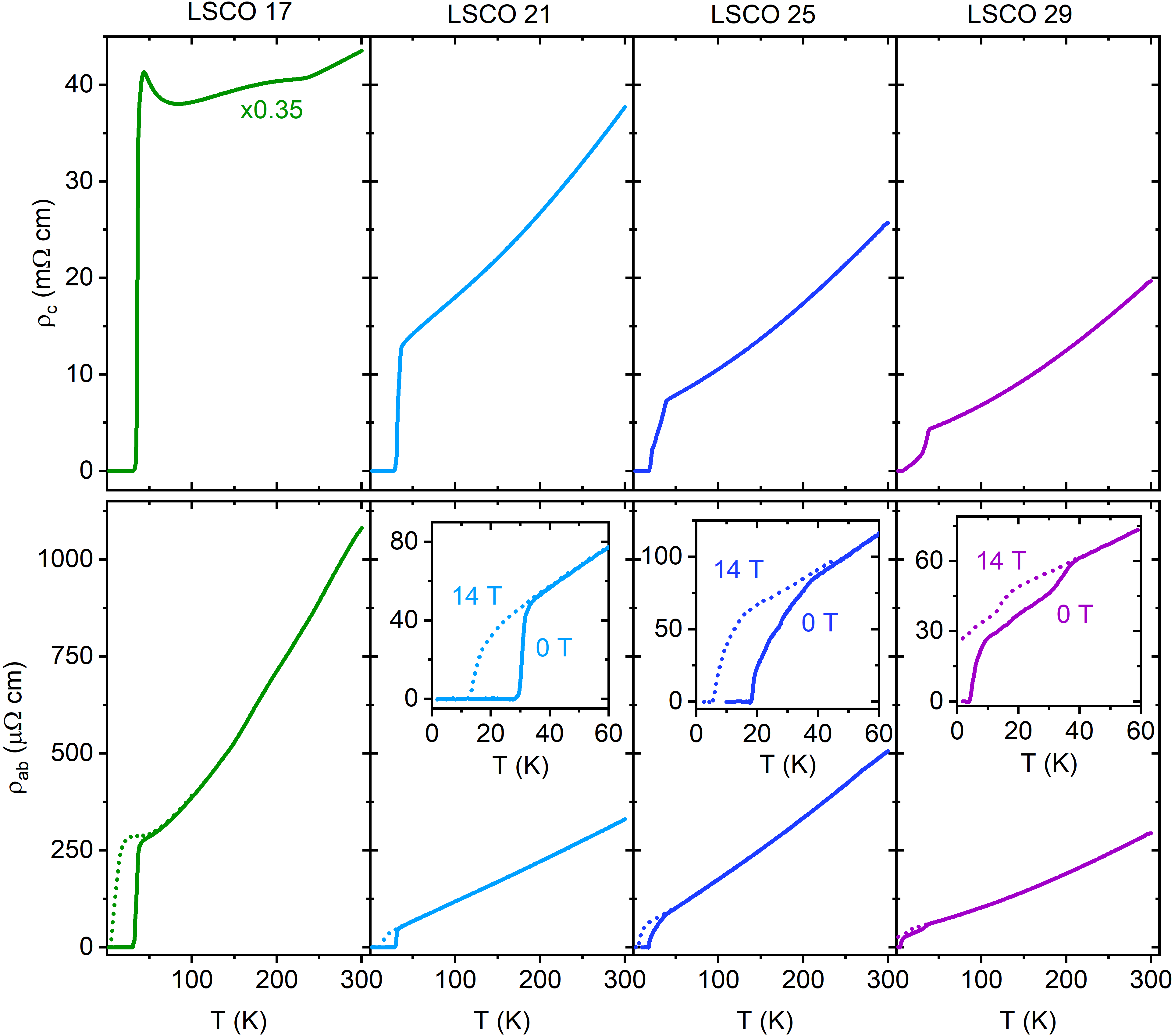}
    \caption{\label{fg:rho}  Resistivity vs.\ $T$ measured along the $c$ axis (top row) and $a$ axis (bottom row) for \lsco\ with $x=0.17$, 0.21, 0.25, and 0.29, as labelled at the top. In the first panel, $\rho_c$ for $x=0.17$ has been multiplied by 0.35 to fit it into the scale relevant for higher doping.  In the insets, dotted lines show $\rho_{ab}$ measured in the presence of a 14-T magnetic field applied along the $c$ axis. }
\end{figure*}

{\newr We note that superconductivity at $p=0.29$ appears to be inconsistent with the single-crystal results of Cooper {\it et al.} \cite{coop09} and Hussey {\it et al.} \cite{huss11}.  From the Supporting Online Material to Ref.~\onlinecite{coop09}, the value of $p$ for those samples was determined from $T_c$ assuming the formula for $T_c(p)$ originally proposed by Tallon and coworkers \cite{pres91}; that formula has also been utilized to determine $p$ for the thin films studied in \cite{bozo16}.  According to that formula, $T_c$ goes to zero at $p=0.27$.   In contrast, our experimental results for superconductivity beyond $p=0.27$ are consistent with the single-crystal results of Yamada and coworkers \cite{ikeu03}, as indicated in Fig.~1 of Ref.~\onlinecite{li22}. Again, for $p>p^\ast$, we have previously presented the case that the superconducting state is inhomogeneous due to intrinsic disorder of the Sr dopants and poor Coulomb screening \cite{li22}.  Such behavior is consistent with the model to be discussed in Sec.~\ref{sc-model}.}

\subsection{Anisotropic Resistivity}

 In Fig.~\ref{fg:rho}, we compare the temperature dependence of  $\rho_c$ (top) and $\rho_{ab}$ (bottom) for four compositions of LSCO.  Let us first consider the behavior of $\rho_c$.  For $x=0.17$, $\rho_c$ is large in magnitude and there is a kink below 250~K corresponding with the structural transition from tetragonal to orthorhombic symmetry; this is followed by an upturn on cooling below 80~K, suggestive of insulating behavior (consistent with the results reported in high magnetic fields \cite{boeb96,capr20}), before reaching the onset of superconductivity.    In comparison, the results for $x=0.21$, 0.25, and 0.29 show a reduced (though still large) magnitude and a metallic temperature dependence in the normal state. 
 
\begin{figure*}[t]
 \centering
    \includegraphics[width=1.6\columnwidth]{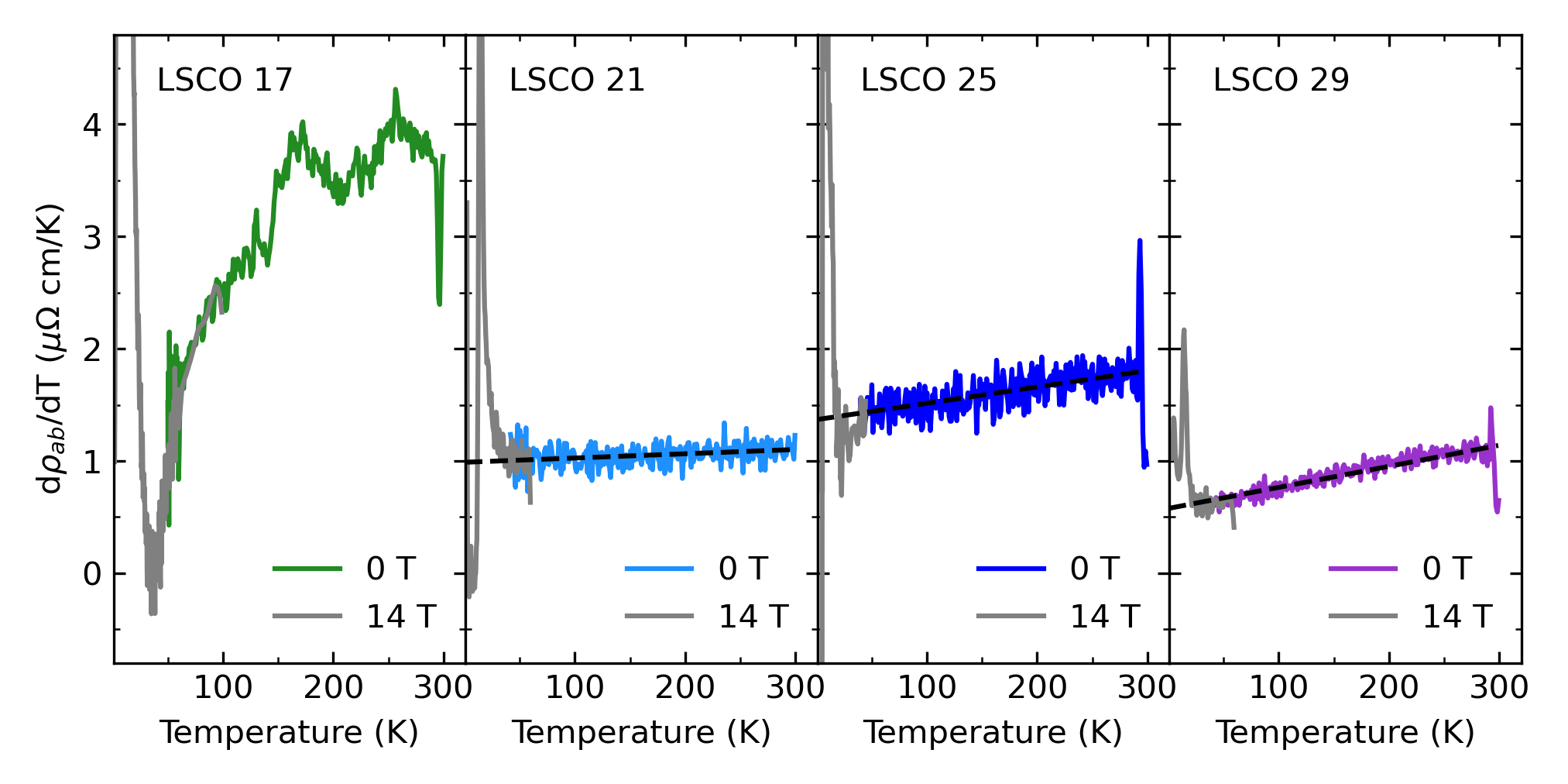}
    \caption{\label{fg:drho} Derivative of $\rho_{ab}$ with respect to temperature for the \lsco\ samples with $x=0.17$, 0.21, 0.25, 0.29.  The zero-field results are shown down to 50~K for $x=0.17$ and to 40~K for the others.  A 3-point triangular smoothing window has been applied; the maximum temperature step is 1~K. }
\end{figure*}

Next we consider $\rho_{ab}$.  For $x=0.17$, we see a metallic temperature dependence; on depressing the superconducting transition by applying a 14~T magnetic field along the $c$ axis,  there is a flattening of $\rho_{ab}$ in the extended normal-state region, which has been associated with the slowing of magnetic fluctuations \cite{bour19} as expected for $p<p^*$.  This behavior contrasts with that of the crystals with $p>p^*$, where we see that $\rho_{ab}$ has reduced magnitude and $T$-dependence.  The insets indicate that depressing $T_c$ with a magnetic field results in a $T$-linear extension of the normal-state behavior to lower temperature.  

To emphasize the strange-metal character of $\rho_{ab}$ for $p>p^\ast$, we plot $d\rho_{ab}/dT$ in Fig.~\ref{fg:drho} .  If $\rho_{ab}$ varies linearly with $T$, then $d\rho_{ab}/dT$ should be a constant, which is approximately the case for $x=0.21$.  There is some $T$-dependent slope that appears for $x=0.25$ and 0.29, but this is small compared to the case of $x=0.17$.  Our results are consistent with a large deviation from Fermi-liquid  behavior in LSCO for $p^*<p<p_c$, as previously identified \cite{coop09,ayre21}.  

Altogether, our transport results support the picture of a transition from non-metal to strange metal in LSCO at $p\sim p^\ast$.  There is a clear crossover in the behavior of $\rho_c$ from low-temperature insulator-like to metallic character.  In $\rho_{ab}$, there is a large jump in the residual resistivity across $p^\ast$; for $x=0.17$, the residual resistivity appears to be $> 250$~$\mu\Omega$-cm, which is above the level at which saturation typically occurs in metals with strong electron-phonon interactions \cite{mooi73,gunn03,huss04}, as discussed in the Introduction.  In particular, our results are consistent with the thin-film data \cite{capr20} shown in Fig.~\ref{fg:rsh}.

\section{Interpretation}
\label{sc_int}

We have outlined our interpretation of the doping-dependent behavior of LSCO in the introduction.  In this section, we present more supporting details, reviewing both experimental results and theoretical analyses.

\subsection{Spin and charge stripe correlations in LSCO}

The spin stripe correlations in LSCO evolve with doping directly from the antiferromagnetic parent phase at $x=0$ \cite{yama98a,birg06}.  For $x>0.05$, the spin stripes are approximately parallel to the Cu-O bonds, and they are detected by neutron scattering at peak positions separated from the antiferromagnetic wave vector in a direction parallel to a Cu-O bond by an amount $\delta a^*$, where $a^*=2\pi/a$ and $a \sim 3.8$~\AA\ is the lattice parameter of a square CuO$_2$ plane.  The doping dependence of $\delta$ \cite{yama98a,zhu23} is indicated by green circles in Fig.~\ref{fg:model}(b).

Most of the magnetic spectral weight is in the inelastic response.  By integrating the magnetic dynamical structure factor over excitation energy and momentum{\newr, as well as any magnetic elastic scattering}, one obtains a quantity proportional to $\langle m^2\rangle$, where $m$ is the {\newr instantaneous} magnetic moment per Cu site.  Several studies \cite{hayd96a,fuji12a,zhu23,waki07b} have provided results that collectively demonstrate the decrease of $\langle m^2\rangle_x$ for doping $x$ relative to the result at $x=0$, as shown by the open green circles in Fig.~\ref{fg:model}(a).  (For samples at larger $x$, we have used available data with $\hbar\omega<100$~meV that have been compared to lower dopings \cite{zhu23,waki07b}.)  We are not aware of a comparable measure of the charge-stripe response vs.\ doping.

Charge stripe order in LSCO was first detected near $x=0.12$ by hard x-ray scattering \cite{crof14}.  Recent resonant soft x-ray scattering studies \cite{wen19,lin20,miao21} have extend the range over which charge-stripe correlations have been observed.  The associated scattering peaks appear at wave vectors of $2\delta a^*$ about reciprocal lattice vectors; the experimental values are presented as blue diamonds in Fig.~\ref{fg:model}(b).  Within the experimental uncertainties (not shown), these charge and spin modulations (measured at low temperature) are mutually commensurate.
The concept that spin and charge stripes are part of the same intertwined state, representative of strong correlations, is supported by numerical studies of the Hubbard and $t$-$J$ models using parameters relevant to hole-doped cuprates \cite{whit98a,zhen17,jian21,jian21c,pons23}.  Calculations yield dynamic \cite{mai22}, as well as static, stripes.

The spin stripe correlations are purely dynamic for $x\gtrsim0.13$ \cite{khay05,chan08}.  Similarly, the charge stripe correlations are also likely to be fluctuating in that range \cite{vona23,bado16}.  Even where stripe order is detected, it can be inhomogeneous.  Measurements with muon spin rotation ($\mu$SR) spectroscopy \cite{savi02,savi05} and $^{63}$Cu \cite{imai17} and $^{139}$La \cite{arse20} nuclear magnetic resonance (NMR) studies indicate that the charge and spin orders occur in only a fraction of the sample volume for $x\sim0.12$.   The key issue is not whether the stripes are static, but the relative area within a CuO$_2$ plane that they occupy.  

In the Appendix, we comment on some experimental results for LSCO thin films with $x>0.3$ that have created some confusion in the field.

\subsection{Model of stripe-to-metal transition in the presence of disorder}
\label{sc-model}

\begin{figure}[t]
 \centering
    \includegraphics[width=\columnwidth]{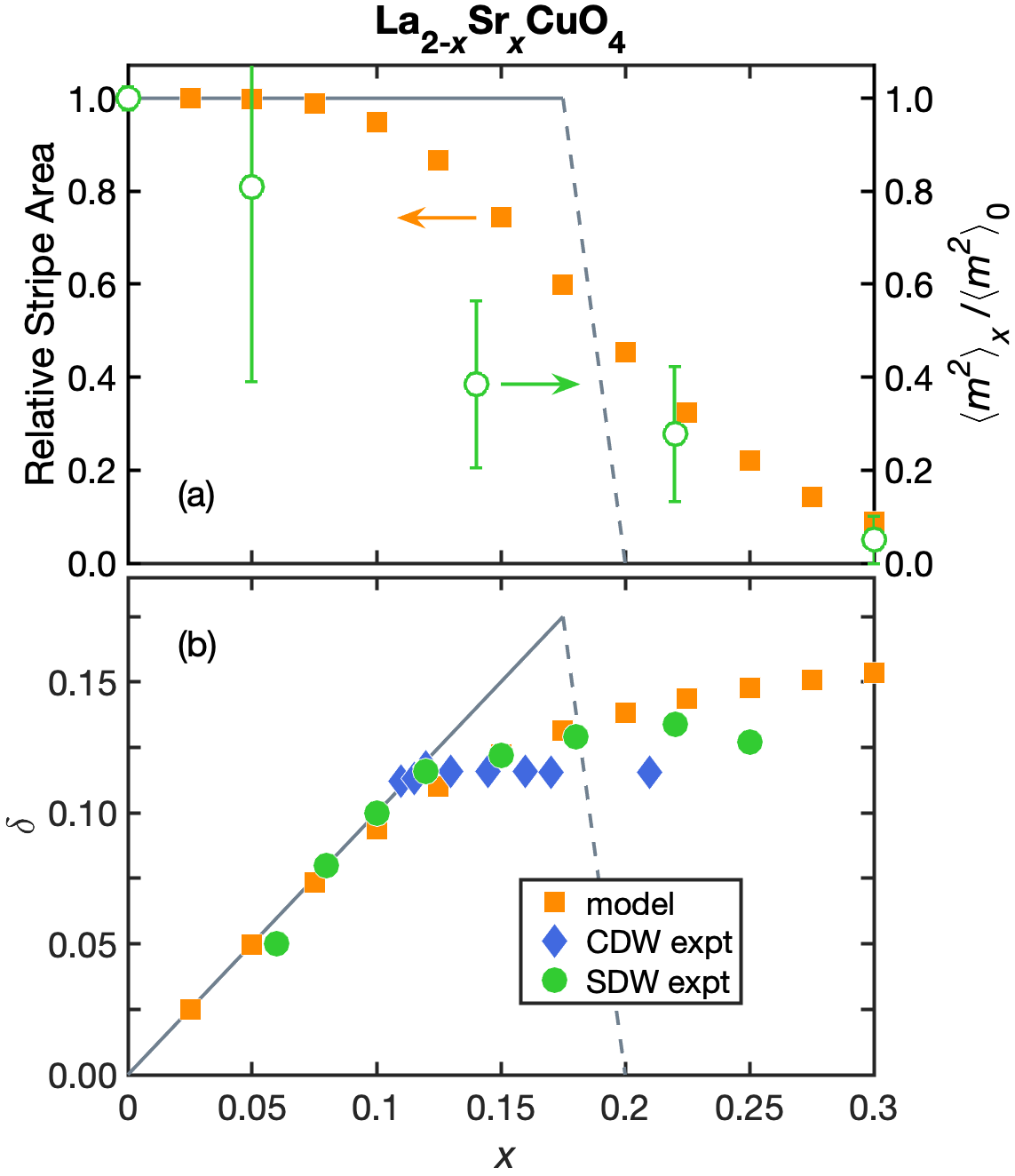}
    \caption{\label{fg:model} (a) Relative stripe area (left axis) for the model with no disorder (line, with step interpolated by a dashed line) and after applying disorder (orange squares) as discussed in the text.  Right axis shows integrated magnetic signal from neutron scattering (green open circles) \cite{hayd96a,fuji12a,zhu23,waki07b}, normalized to $x=0$.  (b) Incommensurability $\delta$ from low--energy (2--3 meV) SDW fluctuations measured by neutron scattering \cite{yama98a,zhu23} (green circles) and from CDW correlations measured with resonant soft x-ray scattering \cite{crof14,wen19,lin20} (blue diamonds), where the measured peak position corresponds to $2\delta$.  These are compared with the model for no disorder (line) and including disorder (orange squares).}
\end{figure}

The stripe correlations in LSCO appear to evolve smoothly with doping, with a continuous decrease in $\langle m^2\rangle$.  To model {\newr this behavior in terms of} a first-order transition from a stripe non-metal phase to a metallic phase, we {\newr have to take account of} disorder.  In the absence of disorder, we assume that stripe correlations with $\delta=x$ are present throughout the CuO$_2$ planes for {\newr $p\le p^\ast$} and are replaced by conventional metal for {\newr $p>p^\ast$}.  {\newr To take account of disorder, as discussed below, we will work with $p$ incremented in steps of 0.025; as such, the critical doping occurs between the grid values of 0.175 and 0.2.  With this constraint, as indicated in Fig.~\ref{fg:model}(b), the model for $\delta$ in the absence of disorder increases linearly up to $x_0 = p_0 =0.175$ (solid line) and then drops to zero at $x=0.2$ (dashed line).}

One of the earliest analyses of the role of intrinsic disorder was performed by Imai and coworkers \cite{sing02a} to understand the distribution of $^{63}$Cu nuclear spin-lattice relaxation rates measured in LSCO by nuclear quadrupole resonance.  The impact of the random distribution of Sr on the La site depends on the volume over which it is averaged.  They obtained a good fit to their data by averaging over a circular area with a radius of 3~nm.  In a later analysis of $^{17}$O nuclear magnetic resonance measurements on LSCO \cite{sing05}, they found a patch radius of 2--5~nm gave a consistent description of the data.

{\newr We assume that the effective local doped-hole concentration $p_{\rm loc}$ varies spatially, but averaging over a sample yields the average value of $p=\langle p_{\rm loc}\rangle$.  To calculate average values of properties that depend on $p$, we need to evaluate that probability that a given value of $p_{\rm loc}$ occurs in the sample.  Assuming that the Sr distribution is random, we expect the probabilities to follow the Poisson distribution \cite{pois24}.  The challenge is that the Poisson distribution depends on the size of the region over which one needs to evaluate.  For a very large characteristic volume, the probability distribution for $p_{\rm loc}$ becomes narrow; as the size of the relevant region decreases, the distribution of possible values broadens and becomes asymmetric. }

The relevant volume to consider is that over which interactions (including long-range Coulomb) determine the local electronic response.
A plausible choice is the coherence volume of the superconducting state. In Ref.~\onlinecite{li22}, we estimated that the coherence volume near optimal doping (with coherence length $\xi_{\rm SC}\sim 1$~nm \cite{hoff02})  corresponds to $N=40$ formula units, involving a circular area within a CuO$_2$ plane and extending over one unit cell along the $c$-axis.  If there are $k$ Sr atoms within a particular local volume, this corresponds to a local doped-hole concentration {\newr $p_{\rm loc} = k/N$}; the average number of Sr atoms in a unit volume is then {\newr $\lambda=\langle p_{\rm loc}\rangle N$}.  Given an integer choice of $\lambda$, the probability $P(k)$ that there are $k$ Sr atoms in a volume is \cite{pois24}:
\begin{equation}
  P(k) = {\lambda^k e^{-\lambda} \over k!}.
\end{equation}
With $N=40$, our choice of {\newr $p_0=0.175$} corresponds to $k_0=7$.  For each value of {\newr $p = \langle p_{\rm loc}\rangle=\lambda/N$}, we can calculate
\begin{equation}
  \langle\delta\rangle = \sum_{k=1}^{k_0}P(k){k\over N},
\end{equation}
{\newr where locally $\delta = p_{\rm loc} = k/N$,}
and
\begin{equation}
  A = \sum_{k=1}^{k_0}P(k),
\end{equation}
where $A$ is the relative stripe area (or volume) for the sample.

For our choice of $N=40$, the relative stripe area and $\langle\delta\rangle$ are plotted as orange squares in Fig.~\ref{fg:model}(a) and (b), respectively.  The model provides qualitative agreement with the decrease in $\langle m^2\rangle_x$ with doping, and it does a good job of describing the smooth saturation of $\delta$ vs.~$x$ at larger $x$.  We note that the relative stripe area reaches a value of $0.5$ for {\newr $p\approx p^\ast$}, and the doping corresponding to maximum $T_c$ is associated with a stripe area of $\sim 2/3$.  

For striped regions to determine the normal-state transport properties, they should percolate across the CuO$_2$ planes.  Given the quasi-2D {\newr character} of the electronic correlations, we expect that the stripe percolation threshold should correspond to $A \approx 0.5$.  Within our picture, it follows that $p^\ast$ should correspond to the stripe percolation threshold. 

Our model assumes the presence of inhomogeneity.  We already noted that NMR studies \cite{sing02a,sing05} provided evidence of electronic heterogeneity in LSCO.  Evidence has also been provided by observations of granular superconductivity in overdoped LSCO \cite{li22}.  In that regime, metallic regions far from a stripe grain would not be superconducting.  Consistent with that, specific heat measurements show that the density of residual, unpaired electronic states grows rapidly for $x\gtrsim0.21$ \cite{wang07}.

For evidence of a mixture of striped and metallic regions, we have to consider $x<p^\ast$ and perturb the system to pin the stripe regions.  One way to do this is to change the symmetry of the lattice structure, as occurs when changing the dopants from Sr to Ba \cite{huck11}.  Zero-field $\mu$SR measurements \cite{gugu16} on \lbco\ with $0.11\le x\le0.17$ found that the magnetic volume fraction (associated with spin stripe order) was $>80$\%\ for $x\le0.15$ but rapidly fell to 50\%\ for $x=0.155$ and below that for 0.17; at the same time, there was no significant change in the local hyperfine field, which probes the spin-stripe order parameter within the ordered volume fraction.

\subsection{Evidence for the stripe percolation threshold}

The role of percolation has been considered previously in terms of the intrinsic charge inhomogeneity, the associated variation in local pairing gap, and the development of superconducting phase order as a function of temperature \cite{miha02,pelc18}.  Our invocation of percolation in the present case is slighlty different, as we are concerned with the low-temperature crossover as a function of doping between local grains with stripe correlations that are associated with strong pairing and those of conventional metal in which intrinsic pairing is weak.

In the study of the percolation of antiferromagnetic order in an insulating compound where the magnetic ions are diluted by substitution of nonmagnetic ions, it is possible to determine the percolation threshold through measurements of the ordered magnetic moment and the spin-spin correlation length.  A good example is the study of the impact of the substitution of nonmagnetic Zn and Mg for Cu in \lco\ \cite{vajk02}.  In the present case, this is not practical, as the charge and spin stripe correlations are generally dynamic at low temperature and in the presence of bulk superconducting order.  The spin-spin and charge-charge correlation lengths of dynamic stripes do not provide a useful measure of the evolution of the volume fraction of stripe correlations with doping.

Application of a $c$-axis magnetic field can depress the superconductivity and enhance the strength of spin \cite{khay05,chan08,lake01,lake02,wen08b} and charge \cite{huck13,wen23} stripe orders.  There is a question, however, of how large a magnetic field is needed to suppress all superconducting correlations in LSCO and to induce stripe order.  The onset of the freezing of Cu spins can be detected by  $^{139}$La nuclear magnetic resonance, which can be performed in high fields.  Such measurements, together with high-field ultrasound measurements, indicate that the magnetic field at which the onset of spin freezing is detected rises linearly with $x$ \cite{frac20,vino22}, consistent with an extrapolation of the onset of static spin-stripe order that was detected by neutron scattering \cite{khay05,chan08}.  For $x=0.17$, the onset field for spin freezing is approximately 30~T, where the upper critical field for superconductivity is estimated to be $\sim50$~T \cite{frac20}.  The signature of the spin freezing becomes weak for $x\gtrsim0.17$; ultrasound indicates a weak onset for $x=0.188$, but the signature is absent for samples with $x\ge0.21$.  These results are consistent with our model.  Once the volume of the stripe correlations falls below the percolation value, it should not be possible to induce order even by suppressing competing orders.

\begin{figure}[t]
 \centering
    \includegraphics[width=0.9\columnwidth]{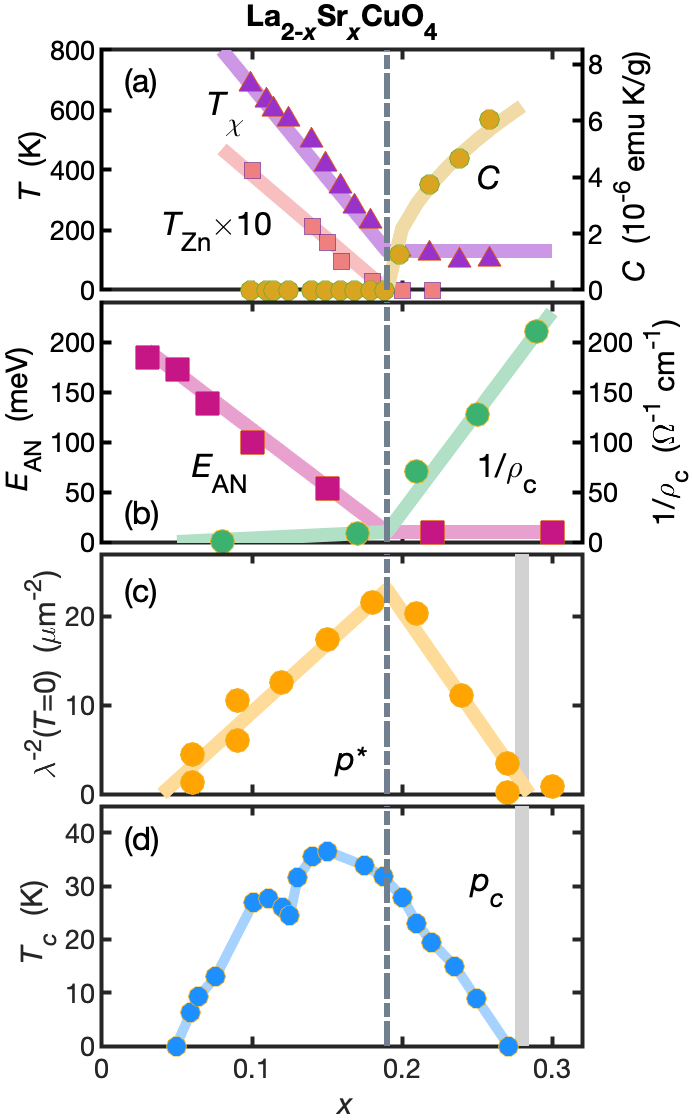}
    \caption{\label{fg:comp}  Comparison of doping dependence of various quantities in LSCO.  (a) Magnetic correlations: $T_{\chi}$ (triangles) is the temperature at which the magnetic susceptibility of polycrystalline samples is maximum, from a fit to 2D AFM response; at large doping, a Curie component (maximum at $T=0$) of magnitude $C$ (circles) dominates \cite{naka94}. $T_{\rm Zn}$ (squares) is the temperature at which spin freezing becomes apparent in $\mu$SR measurements of LSCO doped withe 1\%\ Zn \cite{pana02}. (b) $E_{\rm AN}$ (squares) is the effective antinodal (pseudogap) energy from angle-integrated photoemission measurements \cite{yosh07}; circles indicate $1/\rho_c(T=50\,{\rm K} )$ \cite{li22}.  (c) Inverse square of the magnetic penetration depth in the low $T$ limit (circles), which is proportional to the superfluid density \cite{lemb11}.  (d) $T_c$ determined by the midpoint of the Meissner magnetization \cite{taka89b}.   The vertical dashed line (solid line) indicates $p^\ast$ ($p_c$); solid lines emphasize the trends of the data with doping.}
\end{figure}

It is not practical to perform all measurements at such high magnetic fields.  Instead, we consider practical low-field measures of the doping dependence of spin correlations.  One is provided by the magnetic susceptibility.  For isolated spins, the susceptibility follows the Curie behavior, growing with cooling as $1/T$; however, for a 2D network of antiferromagnetically-coupled spins, the susceptibility decreases as correlations grow beyond nearest neighbors \cite{huck08}.  The magnetic susceptibility data measured on polycrystalline samples of LSCO show, in the underdoped regime, a peak at a temperature $T_{\chi}$, with a decrease at $T<T_{\chi}$ \cite{naka94}, as shown in Fig.~\ref{fg:comp}(a).  It is observed that $T_{\chi}$ drops as $p\rightarrow p^*$, consistent with the approach of an antiferromagnetic system to a percolation limit. Beyond that point, a Curie-like component develops \cite{naka94,kais12}, as one would expect in a system in which antiferromagnetic correlations are limited to finite grains \cite{bree73}.

Another probe of the spin-stripe correlations is provided by substitution of a small fraction (1\%) of Zn for Cu \footnote{Here, we are interested in a small perturbation of the planes that does not disrupt the intrinsic correlations; we explain in the Appendix why we believe that larger Zn concentrations, as studied in Ref.~\cite{risd08,adac08}, which induce magnetic order for $x\gg p^\ast$ are not relevant to this point.}, where it has been shown that a small concentration ($\lesssim1$\%) of Zn  has a similar effect on the superconductivity as a $c$-axis magnetic field \cite{nach96,loza21a}.  Neutron scattering measurements have demonstrated that Zn-doping enhances spin-stripe order \cite{kimu99,wen12a}.  The temperature $T_{\rm Zn}$ at which the spin correlations begin to freeze has been measured as a function of doping by $\mu$SR \cite{pana02}; as one can see in Fig.~\ref{fg:comp}(a), $T_{\rm Zn}$ drops toward zero at $p^*$.  (Related results have been reported for Zn-doped \ybco\ \cite{mend99,tall01}.)  This is consistent with a picture in which dynamic correlations that percolate across the sample can be pinned by Zn defects, but the pinning is no longer effective beyond the percolation threshold.

Local antiferromagnetic spin correlations associated with spin stripes strongly scatter the electronic excitations with wave vectors near $(\pi/a,0)$ and $(0,\pi/a)$ \cite{wu22,krie22b} that correspond to the ``antinodal'' (AN) region, where the superconducting $d$-wave gap should have its maximum.  The resulting damping causes a shift in the effective peak energy, $E_{\rm AN}$, detected by photoemission in this region \cite{ino98,sato99,yosh07}; $E_{\rm AN}$ decreases with doping, as shown in Fig.~\ref{fg:comp}(b), in a fashion that correlates with the approach to the stripe percolation limit.  The appearance of coherent AN states for $p>p^*$ has been observed in \bscco\ by scanning tunneling microscopy (STM) \cite{fuji14a} and angle-resolved photoemission spectroscopy (ARPES) \cite{droz18,chen19a}.  In LSCO, electronic dispersion for wave vectors along the $c$ axis is observed only for $p>p^*$ and only with an AN wave vector component \cite{hori18}. 

The electronic states in the AN region have a relatively flat dispersion, which, in combination with being near the Brillouin zone boundary, means that they dominate the density of states near the Fermi level \cite{zhon22}.  This also makes the AN states and their degree of coherence important for conductivity between planes, along the $c$ axis.  We have already seen in Fig.~\ref{fg:rho} that the temperature dependence of $\rho_c$ crosses from insulator-like to metallic as $x$ crosses $p^\ast$.  As a measure of the $c$-axis conductivity, we plot $1/\rho_c$, measured in the normal state at 50~K, in Fig.~1(b) \cite{li22}.  The gradual rise of the metallic conductivity in the overdoped region is distinct from the variation of the density of states at the Fermi energy, which has a maximum at $p\sim0.21$ due to a Lifshitz transition \cite{zhon22}.

To provide context, we summarize in Fig.~\ref{fg:comp} experimental results for several different doping-dependent properties measured on LSCO.  Figure~\ref{fg:comp}(c,d) show that a measure of the superfluid density peaks at $p\approx p^*$, which is higher than the doping at which the superconducting transition temperature, $T_c$, peaks.  The quantity plotted in Fig.~\ref{fg:comp}(c) is the inverse square of the magnetic penetration depth (at $T\ll T_c$) measured by mutual inductance on LSCO films \cite{lemb11}.  The data shown are consistent with a collection of related measures of superfluid density in LSCO summarized in Ref.~\cite{rour11}, as well as with results from $\mu$SR on other cuprates \cite{bern01}.
(The only exception involves the encapsulated LSCO thin films studied by Bozovic {\it et al.} \cite{bozo16}, where a continuous decrease of superfluid density was observed for $p>0.16$.)

\section{Discussion}
\label{sc_dis}

\subsection{First-order transition}

We have presented the case for a first-order transition from correlated stripes to conventional metal that is masked by disorder.
Such a picture is compatible with a recent study of in-plane optical conductivity in overdoped LSCO \cite{mich21}, where a coherent Drude component was found to grow monotonically with $p$, while the incoherent mid-infrared (MIR) component decreases in the overdoped regime, correlates with $T_c$, and is absent for $p\gtrsim p_c$.  We associate the MIR component with the regions of surviving stripe correlations; it has previously been shown that the energy scale of the MIR conductivity in cuprates matches well with that observed in two-magnon Raman scattering \cite{tran21a,suga03}, which evolves continuously from the parent antiferromagnetic insulator. 

Angle-resolved photoemission measurements on LSCO \cite{hori18,zhon22} suggest that the nominal Fermi surface undergoes a Lifshitz transition, from a hole-like to electron-like pocket, at $x\sim 0.21$.  If the system were electronically homogeneous, one would expect a corresponding change in sign of the Hall coefficient from positive to negative; however, the sign change does not occur until $p>0.3$ \cite{taka89b}.  We suggest that the intrinsic disorder we have discussed can explain this effect.  

The model commonly used to describe cuprates is the Hubbard model \cite{arov22,qin22}.  It typically excludes the extended Coulomb interaction and limits the number of atomic orbitals per unit cell, most often to one; nevertheless, even the simplified model is challenging to evaluate, especially at low temperature. There is no consensus on a phase diagram as a function of doping for the case of intermediate coupling, relevant to cuprates, where the bandwidth is comparable to the onsite Coulomb repulsion, but there are notable results.   

The temperature-dependent crossover behavior in properties such as magnetic susceptibility and $\rho_c$ associated with the onset of the pseudogap phase has been associated with the Widom line \cite{xu05w}, which terminates at zero temperature in a first-order transition \cite{sord13}.   The value of $p$ at the transition is sensitive to model parameters \cite{wu22,wals23}, but is roughly consistent with the experimental $p^\ast$.  It is recognized that the antinodal pseudogap is caused by antiferromagnetic spin fluctuations, as calculated with the dynamical cluster approximation (DCA) to dynamical mean-field theory \cite{gunn15,wu22}; in terms of spatial correlations, the spin fluctuations are associated with stripe correlations obtained in other calculations such as density-matrix renormalization group \cite{whit98a} and related approaches \cite{zhen17}, including tensor-network calculations \cite{pons23}, as well as with DCA usig larger cluster sizes \cite{mai22}.   These theoretical results are compatible with our empirical picture.

\subsection{Strange-metal behavior in LSCO}

A common approach has been to associate strange-metal behavior with quantum critical fluctuations of some order parameter.  We have presented empirical evidence, both our own and from the literature, that the change at $p^\ast$ is associated with a first-order transition that is masked by disorder.  If we are correct, then the quantum-critical analyses are not directly relevant.   Nevertheless, models that consider antiferromagnetic fluctuations \cite{teix23} may have partial relevance.  Low-energy incommensurate spin fluctuations are certainly present at $p>p^\ast$ \cite{zhu23}, and the weight at low frequency grows on cooling toward $T_c$ \cite{waki04}; the difference is that they cannot diverge at $\hbar\omega=0$ as $T\rightarrow 0$ (when superconductivity is suppressed), because, we argue, they are confined to finite bubbles.

Patel {\it et al.} \cite{pate23} have evaluated a model with several possible contributions to electronic scattering; a $T$-linear scattering rate is given by a component describing spatially-random interactions.  The nature of the interaction is not specified, but the disorder could be consistent with our evidence for heterogeneity.

Recent calculations within the Hubbard model using determinantal quantum Monte Carlo calculations show promising results \cite{huan19}, with $T$-linear resistivity for $p\gtrsim0.2$; however, they are limited to temperatures far above room temperature.  Perhaps more significant are calculations using the dynamical cluster approximation that show, just beyond the first-order termination of the strongly-correlated pseudogap phase, a non-Fermi-liquid phase with a scattering rate that varies linearly with $T$ \cite{wu22}; here, the calculations extend down to $T$ of order 100~K.  Importantly, it is found that, in this phase, the main contribution to the electronic self-energy is from antiferromagnetic spin fluctuations \cite{wu22}.  These calculations do not explicitly include disorder, which would seem to be different from our picture.  On the other hand, an effective model of electrons scattering from fluctuating impurity magnetic moments yields the desired behavior \cite{ciuc23}.  We would view the impurity moments as representative of stripe patches.

Some theorists have emphasized the role of defect scattering and have proposed that it is responsible for the decrease in $T_c$ with overdoping \cite{leeh17,ozde22,pal23}.  The impact of defect scattering on quasiparticles shows up in the residual resistivity, $\rho_0$.  We agree that there is substantial disorder from dopants, and we have emphasized its impact on the evolution of correlations with doping in LSCO.  We also acknowledge that there is a significant residual resistivity in overdoped samples.  A linear extrapolation of $\rho_{ab}(T)$ for our $x=0.21$ crystal yields $\rho_0\approx15$~$\mu\Omega\,$cm, compatible with high-field studies \cite{coop09,boeb96}.  Dividing by the interlayer spacing (6.6~\AA) to convert to sheet resistance, we find that $R_0\approx 10^{-2}\, h/e^2$.  For comparison, $R_0$ for an excellent 2D conductor, PdCoO$_2$, is $5\times 10^{-6}\, h/e^2$ \cite{hick12}.  

Where we differ is on the role of defect scattering as the controlling factor in limiting $T_c$ in overdoped samples.  The analysis of dirty $d$-wave superconductivity is based on weak-coupling BCS theory.  We argue that weak-coupling BCS theory is not relevant to describing superconductivity in the cuprates.  As should be clear from Fig.~\ref{fg:comp}, $T_c$ is largest for $x<p^\ast$, where, as we have discussed, the stripe correlations dominate.  In this regime, the electronic state is non-metallic as indicated by the high-field resistivity results in Fig.~\ref{fg:rsh}; the issue of residual resistivity has little relevance relative to the effects of antiferromagnetic scattering that localize the charge carriers.  Nevertheless, superconductivity develops in this environment.  We have argued that it is the loss of the strongly-correlated environment that leads to the decay of superconductivity in the very overdoped regime \cite{li22}.

\subsection{Other cuprates}

Suppose our description of LSCO is approximately correct; does it have any relevance for other cuprate families?

First consider our interpretation of a first-order transition converted to a crossover at $p^\ast$.  An immediate challenge is the observation of quantum oscillations in \ybco\ \cite{doir07,seba08,seba15}.  The quantum oscillation signals, observed as a function of magnetic field at low temperature in the absence of superconducting order, are interpreted in terms of the Fermi surface of a conventional metal \cite{chak11,seba12}.  Experiments indicate that quantum oscillations are observable in the doping range $0.08\lesssim p \lesssim0.18$ \cite{seba12,rams15}, where the upper limit occurs close to the estimated $p^\ast$ \cite{tall01}.

Recent measurements indicate that the quantum oscillations come uniquely from very small pockets, presumably associated with nodal state, with no contribution coming from antinodal states \cite{hart20}.  This is consistent with measurements of $\rho_c$, which exhibit a low-temperature upturn for $p\lesssim p^\ast$ \cite{ito91,take94,coop00}, demonstrating non-metallic character.  For $p\lesssim 0.10$, insulating character within the CuO$_2$ planes was inferred from thermal conductivity measurements \cite{sun04}.  The CDW features seen in zero field show differences from LSCO \cite{hayd24}; however, antiferromagnetic spin excitations are observed across the full range \cite{dai01,hink10,suga03}, and the incommensurate wave vector of the lowest-energy spin fluctuations evolves with doping in a fashion very similar to that in LSCO \cite{dai01,enok13}.   While the normal-state character of YBCO for $p<p^\ast$ is complicated, it is clear that it is far from a conventional metal.

With respect to the role of inhomogeneity in the strange-metal behavior, we have already noted the heterogeneous superconducting gap seen in STM studies of (Pb,Bi)$_2$Sr$_2$CuO$_{6+\delta}$ \cite{trom23,ye23}.  Linear resistivity in overdoped Bi$_2$Sr$_2$CuO$_{6+\delta}$ has recently been reconfirmed \cite{zang23}.  ``Planckian'' dissipation has also been observed in overdoped \bscco\ \cite{legr19}, where angle-resolved photoemission finds a rapid decay in the antinodal superconducting gap size \cite{vall20} despite the enhanced quasiparticle coherence relative to $p<p^\ast$ \cite{chen19a}; STM shows substantial heterogeneity of local superconducting gap sizes in this system \cite{mcel05a,fang06}.

Tl$_2$Ba$_2$CuO$_{6+\delta}$ (Tl2201) is another system in which a $T$-linear contribution to $\rho_{ab}$ has been reported for overdoped samples \cite{huss13,ayre21}.  Measurements suggest that it is a relatively clean system \footnote{Note that substitutional defects of Cu on the Tl site can occur at the level of 7\%\ \cite{mack93}.}: a full hole-like Fermi surface has been detected by photoemission \cite{plat05} on a crystal with $p\approx0.27$, where the Hall number corresponds to $1+p$ \cite{putz21}.  Quantum oscillations have also been observed for crystals with $p=0.27$ and 0.30 \cite{vign08,bang10}.
In contrast, $\mu$SR measurements \cite{uemu93,nied93} indicated that the ratio of the superfluid density to the effective mass decreases with $T_c$ on the overdoped side, which, assuming that the effective mass does not change, is inconsistent with weak-coupling BCS theory in the clean limit.  While scattering from defects might explain the effect, that would appear to be incompatible with the photoemission and quantum oscillation results.  This led Uemura \cite{uemu01} to argue for phase separation in overdoped cuprates between superconducting and normal-metal phases, which is compatible with our disorder model. 

Nuclear quadrupole resonance measurements in Tl2201 using $^{63}$Cu nuclei provide evidence for antiferromagnetic spin fluctuations  in the normal state that weaken with overdoping, similar to other cuprate families \cite{fuji91,kamb93};  however, the coefficient of the $T$-linear contribution to $\rho_{ab}$, when expressed in terms of sheet resistance, is only a quarter of that in LSCO for the same doping \cite{ayre21}.  Hence, the impact of spin fluctuations is probably weaker than in LSCO, enabling more coherent quasiparticles.

\section{Conclusions}
\label{sc_con}

The in-plane resistivity in LSCO is observed to vary approximately linear with temperature down to low temperature for hole concentrations greater than $p^\ast\sim0.19$ and up to $p_c$, the superconductor-to-metal transition.  The cause appears to be residual, spatially-inhomogeneous antiferromagnetic spin fluctuations associated with residual stripe correlations.  This behavior is different from the $T^2$ dependence expected for a Fermi liquid but modest compared to the non-metallic behavior observed for $p<p^\ast$.

Below $p^\ast$, many properties exhibit the effects of strong correlations, and in LSCO charge and spin stripes are one consequence of these.  A consistent interpretation of a range of experimental results is achieved with the assumption that there is a first-order transition between strongly-correlated and more conventional phases.  The transition appears as a crossover at $p^\ast$ due to the presence of intrinsic disorder of the dopant ions.  The stripe correlations percolate across the CuO$_2$ planes for $p < p^\ast$, while they are limited to finite puddles in the overdoped region.  Without the disorder, we expect that there would be no superconductivity or strange-metal behavior for $p>p^\ast$.

\begin{acknowledgments}
We thank S. A. Kivelson, A. Tsvelik, {\newr and Andreas Weichselbaum} for valuable comments.
Work at Brookhaven is supported by the Office of Basic Energy Sciences, Materials Sciences and Engineering Division, U.S. Department of Energy (DOE) under Contract No.\ DE-SC0012704.   
\end{acknowledgments}

\appendix*
\section{Discussion of other work}

\subsection{Studies of samples at $p > p_c$}

There has been some confusion associated with an early resonant inelastic x-ray scattering study of magnetic excitations in LSCO thin films with $x$ as high as 0.4 \cite{dean13}.  These have been interpreted by some as evidence that antiferromagnetic spin fluctuations are still present beyond the superconducting dome.  There are a number of problems with this interpretation.  An important one is that the measurements do not come sample wave vectors close to the antiferromagnetic wave vector. This is because measurements with x-rays at the energy of the Cu $L_3$ edge cannot provide such a momentum transfer; one can only reach the antiferromagnetic zone boundary.  That means that the measurements are essentially within a ferromagnetic Brillouin zone.  When there is antiferromagnetic order, then the spin-wave dispersion is the same in both zones, though the intensity is very weak in the ferromagnetic zone.  In the absence of order, there is no such symmetry constraint.

The measurements were also performed with a coarse energy resolution.  More recent measurements \cite{meye17,roba19} with better energy resolution have shown that there is a large change in energy width of the excitations with doping and some reduction of intensity near the zone boundary.  Furthermore, theoretical calculations suggest that ferromagnetic excitations may develop at high doping \cite{jia14}, while truly antiferromagnetic fluctuations decrease substantially \cite{huan17}.  

The key point is that measurements by inelastic neutron scattering on an LSCO crystal with $x=0.30$ found that spin fluctuations near the antiferromagnetic wave vector at energies below 100 meV have negligible intensity \cite{waki04,waki07b}.  The RIXS studies do not contradict this result.

A new publication might create confusion in a different direction.  In films with $x=0.35$, 0.45, and 0.6, a diffraction peak at $(\delta,0,L)$ with $\delta\approx1/6$ has been been seen in RIXS measurements, with little intensity change up to 300~K \cite{li23}.  This peak has been interpreted as evidence of charge order.  We feel that such an interpretation is premature and questionable.  In particular, one needs further characterization of these samples and the level of apical-oxygen vacancies, which are known to be an issue for $x\gtrsim0.3$ \cite{rada94}.  A related issue has occurred in studies of NdNiO$_2$ films where incomplete oxygen reduction can result in periodic partial occupancy of apical O sites that has been detected both by electron diffraction and RIXS \cite{raji23,parz24}; the peak in RIXS can be misinterpreted as evidence of charge order.

We have sensitivity to this issue, as one of us (JMT) has his own experience with misinterpreting a set of unexpected superlattice peaks (measured by neutron diffraction in La$_2$NiO$_{4+\delta}$) in terms of CDW order \cite{tran93}.  Further investigation revealed that they were a consequence of the ordering of interstitial oxygens \cite{tran94b}.

\subsection{Impact of larger concentrations of Zn dopants}

In considering the impact of Zn doping on spin freezing, as in the data of Ref.~\onlinecite{pana02} shown in Fig.~\ref{fg:comp}(a), we are interested in considering the Zn as a minimal perturbation that depresses the superconductivity without significantly modifying the intrinsic stripe correlations.  In fact, Zn locally acts as a strong perturbation, preventing the motion of holes through the Zn site, which can limit the flipping of spins on neighboring Cu sites.  From the $\mu$SR study of Nachumi {\it et al.} \cite{nach96}, where the impact of Zn was studied in steps of 0.25\%\ in impurity concentration, 1\%\ Zn is already a significant perturbation.  A study combining $\mu$SR and neutron diffraction showed that larger Zn concentrations can depress spin-stripe and superconducting transitions in a similar fashion \cite{gugu17}. 

Studies by Adachi {\it et al.} \cite{adac08} and Risdiana {\it et al.} \cite{risd08} showed that one can enhance the magnetic volume fraction measured by $\mu$SR at $T<1$~K by adding more Zn.  For example, with 3\%\ Zn in LSCO, it was possible to detect some local magnetism for doping up to $x=0.27$ \cite{risd08}.   We suggest that 3\%\ Zn is a very large perturbation that significantly modifies the intrinsic correlations, and hence those results do not provide a useful measure of the unperturbed volume fraction of stripe correlaions.

\bibliography{LNO,theory}

\end{document}